
\input jytex.tex   
\typesize=10pt
\magnification=1200
\baselineskip=17truept
\hsize=6truein\vsize=8.5truein
\sectionnum=0
\sectionnumstyle{arabic}


\def\eql#1{\eqno\eqnlabel{#1}}
\def\ref{\reference}
\def\peq{\puteqn}
\def\pref{\putref}

\def\mbox#1{{\leavevmode\hbox{#1}}}
\def\hspace#1{{\phantom{\mbox#1}}}

\def\frac#1/#2{\leavevmode\kern.1em
\raise.5ex\hbox{\the\scriptfont0 #1}\kern-.1em/\kern-.15em
\lower.25ex\hbox{\the\scriptfont0 #2}}
\def\sfrac#1/#2{\leavevmode\kern.1em
\raise.5ex\hbox{\the\scriptscriptfont0 #1}\kern-.1em/\kern-.15em
\lower.25ex\hbox{\the\scriptscriptfont0 #2}}

\def\gtorder{\mathrel{\raise.3ex\hbox{$>$}\mkern-14mu
             \lower0.6ex\hbox{$\sim$}}}
\def\ltorder{\mathrel{\raise.3ex\hbox{$<$}|mkern-14mu
             \lower0.6ex\hbox{\sim$}}}

\def\semidirprod{\rlap{\ss C}\raise1pt\hbox{$\mkern.75mu\times$}}

\def\for{\lower6pt\hbox{$\Big|$}}
\def\fish{\kern-.25em{\phantom{abcde}\over \phantom{abcde}}\kern-.25em}

\def\boxit#1{\vbox{\hrule\hbox{\vrule\kern3pt
        \vbox{\kern3pt#1\kern3pt}\kern3pt\vrule}\hrule}}
\def\dalemb#1#2{{\vbox{\hrule height .#2pt
        \hbox{\vrule width.#2pt height#1pt \kern#1pt
                \vrule width.#2pt}
        \hrule height.#2pt}}}


\def\noin{\noindent}

\def\be{\beta}

\def\ep{\epsilon}

\def\th{\theta}

\def\etc{{\it etc. }}

\def\eg{{\it e.g. }}

\def\cf{{\it cf }}
\def\pa{\partial}

\def\gap{\vskip 20truept}

\def\bra{\langle}
\def\ket{\rangle}

\def\3j#1#2#3#4#5#6{\left\lgroup\matrix{#1&#2&#3\cr#4&#5&#6\cr}
\right\rgroup}


\def\anp#1#2#3{{\it Ann. Phys.} {\bf {#1}} (19{#2}) #3}

\def\jpa#1#2#3{{\it J. Phys.} {\bf A{#1}} (19{#2}) #3}

\def\np#1#2#3{{\it Nucl. Phys.} {\bf B{#1}} (19{#2}) #3}
\def\pl#1#2#3{{\it Phys. Lett.} {\bf {#1}} (19{#2}) #3}
\def\prp#1#2#3{{\it Phys. Rep.} {\bf {#1}} (19{#2}) #3}
\def\pr#1#2#3{{\it Phys. Rev.} {\bf {#1}} (19{#2}) #3}

\def\prs#1#2#3{{\it Proc. Roy. Soc.} {\bf A{#1}} (19{#2}) #3}

\vglue 1truein
\rightline {MUTP/94/2}
\gap
\centerline {\bigfonts \bf Remarks on geometric entropy}
\vskip 15truept
\centerline{J.S.Dowker}
\vskip 10 truept
\centerline {\it Department of Theoretical Physics,}
\centerline{\it The University of Manchester, Manchester, England.}
\vskip 40truept
\centerline {Abstract}
\vskip 10truept
The recently discussed notion of geometric entropy is shown to be related to
earlier calculations of thermal effects in Rindler space. The evaluation is
extended to de Sitter space and to a two-dimensional black hole.

\rightline {January 1994}
\vfill\eject
\section{\bf Introduction}

A number of recent works [\pref{{Callan}, {Susskind}, {Kabat}}] have been
concerned with the first quantum correction to the
Bekenstein-Hawking black hole entropy. The calculations involve, among other
things, the statistical mechanics of fields in Rindler space-time as
a simply analysed case that might throw light on the black hole system.
Field theory in Rindler space-time has been extensively investigated over
the past 20 years [\pref{Birrell,Takagi}]. Complete references are not
possible in this note but the fact that tracing over fields restricted to
half Minkowski
space yields a thermal average was early known, \cf [\pref{Israel}], and is
sometimes called the thermalisation theorem [\pref{{Sewell},{Fulling}}].

Because of the infinite red-shift at the horizon, the integrated (total)
thermodynamical quantities diverge. A regularisation can be effected by
integrating up to a finite distance from the horizon, \cf [\pref{t'Hooft}].

As a byproduct of an analysis of quantum field theory around a cosmic
string, the finite-temperature energy density in Rindler space was
determined in [\pref{condef}]. The basic fact is that the conical part of
the cosmic string metric is a euclideanised part of Rindler space-time, \cf
[\pref{qftcone}]. In this paper we wish to investigate the
relevance of this calculation and of an earlier one [\pref{TGR}].

\section{\bf Some thermal averages in Rindler space-time.}

We write the Rindler metric (in four dimensions) as
$$
ds^2=Z^2dv^2-dZ^2-dx^2-dy^2
\eql{Rindler}$$
which is compared with the conical cosmic string metric
$$
ds^2=dt^2-dr^2-r^2d\phi^2-dz^2
\eql{cstring}$$ giving the identifications
$\phi\approx-iv$, $r\approx Z$, $t\approx ix$, $z\approx y$. One then has the
equality
$$\bra T^0_0\ket_{\rm Rindler}=\bra T_2^2\ket_{\rm string}.
$$
The periodicity of $\be$ in the angle $\phi$ of the string translates into
a temperature of $1/\be$ in Rindler space and the finite-temperature
expressions in the Rindler case can be read off from the zero temperature
cosmic string results which have been derived over the years by various
means.

The details of our evaluation are in the cited references.
The basic idea is to reperiodise the polar angle from $2\pi$ to $\be$ in
the Green function by a complex contour method. That is to say, we introduce
a conical singularity into the manifold.

A point that should be made is that the vacuum averages were rendered
finite in [\pref{condef}] by subtraction of the Minkowski expressions.
This can be done at the local heat-kernel or Green function level.

We find for $\bra T_0^0\ket$ the expressions
$${\pi^2 T^4\over30}-{1\over480\pi^2Z^4}$$
for spin zero
$${7\pi^2T^4\over120}+{T^2\over48Z^2}-{17\over1920\pi^2Z^4}\eql{endens}$$
for spin $1/2$ and
$${\pi^2T^4\over15}+{T^2\over6Z^2}-{11\over240\pi^2Z^4}$$
for spin 1. All fields are massless.

$T$ is the local Tolman temperature $T=T_0/Z=1/\be Z$. The zero temperature
values were evaluated by Candelas and Deutsch [\pref{Cand}] using the same
Minkowski subtraction. When $\be=2\pi$, the Rindler finite-temperature Green
function
equals the standard Minkowski one and the thermal averages are zero. The
usual statement is that the Minkowski vacuum is a thermal Rindler state
at temperature $T_0=1/2\pi$.

The thermal averages diverge as the horizon is approached just as the
ordinary Casimir vacuum averages diverge when a spatial boundary is
approached. The quick way of seeing this is when the method of images
applies to the construction of the Green function. All this is well known. Its
relevance regarding the calculations in
[\pref{{Callan},{Susskind},{Kabat}}] depends on the significance of the
Minkowski subtraction.

For the time being, we assume that the global quantities are
obtained by integration of the local ones. Thus,
$$E(\be)=\int\bra T_0^0\ket  Z dZ dx dy.$$

Integrating over the range $\ep<Z<\infty$, and letting $x$ and $y$ run over
a region of area $A$, we see, trivially, that $E(\be)$ will be proportional
to $A$ and will diverge as $1/\ep^2$. For spin zero for example
$$
E(\be,\ep,A)={A\pi^2\over60\ep^2\be^4}-{A\over960\pi^2\ep^2}
={A\pi^2\over60\ep^2\be^4}+E(\infty,\ep,A)
$$
and of course this vanishes when $\be=2\pi$, corresponding to the Minkowski
subtraction.

The $T$-dependent terms in (\peq{endens}) are the finite-temperature
corrections to the zero temperature quantities. We shall denote such
quantities by a prime, \eg $\bra T_0^0\ket'$ and $E'(\be)$ and so on.

To obtain the entropy, we need the free energy, related to $E$ by
$$
E={\pa\over\pa\be}\big(\be F\big)
$$
so that, quite generally,
$$ \be F(\be)=\int^\be E(\be) d\be+ C
=\int^\be E'(\be) d\be+ \be E(\infty)+C
$$
where $C$ is a temperature independent constant. Then
$$S=\be E'-\int^\be E'(\be) d\be-C
\eql{entropy1}$$
For spin zero
$$
S(\be,\ep,A)
=\be E'(\be,\ep,A) +{A\pi^2\over180\ep^2\be^3}-C=
{A\pi^2\over45\ep^2\be^3}-C.
$$
Therefore
$$
S(2\pi,\ep,A)
={A\over360\ep^2\pi}-C
\eql{entropy2}$$
The significance of the Minkowski subtraction must now be fairly faced
since it means that at $\be=2\pi$ all thermodynamical quantities like $E$,
$F$ and $S$ vanish, the constant $C$ being adjusted to make this so.

The finite-temperature corrections would be the only terms
obtained if, instead of removing the Minkowski Green function, the zero
temperature ($\be=\infty$) one were subtracted. This was the procedure
adopted in [\pref{TGR}]. Then we would have for $\bra T^0_0\ket'$
$${\pi^2 T^4\over30}$$
for spin zero
$${7\pi^2T^4\over120}+{T^2\over48Z^2}\eql{endens2}$$
for spin $1/2$ and
$${\pi^2T^4\over15}+{T^2\over6Z^2}$$
for spin 1.

This corresponds to setting the constant $C$ equal to zero so that the zero
temperature entropy vanishes. The resulting thermodynamic quantities are
those that would arise by calculating the partition function in the standard
sum-over-states way. Equation (\peq{entropy2}) agrees with the evaluations in
[\pref{{Callan},{Susskind},{Kabat}}]. For the record we give the other
spin results,
$$
S'_{1/2}(2\pi,\ep,A)={11A\over720\pi\ep^2}
$$
$$
S'_{1}(2\pi,\ep,A)={4A\over45\pi\ep^2}.
$$

In two dimensions [\pref{TGR}]
$$
\bra T^0_0\ket'={\pi T^2\over6}
\eql{2den}$$
which for $\be=2\pi$ is Davies' result  [\pref{Davies}].
Substitution into (\peq{entropy1}) yields
$$
S'(\be,\ep)={\pi\over3\be}\ln(D/\ep)
\eql{2dess}$$
where $D$ is an upper limit to the distance from the horizon. Therefore
$$
S'(2\pi,\ep)={1\over6}\ln(D/\ep),
$$
which again agrees with the recent calculations.
\section{\bf de Sitter space and a two-dimensional black hole.}
In [\pref{TGR}] de Sitter space was also treated by the same thermal
technique of re-periodisation. The metric can be written to exhibit a
Rindler-like part,
$$
ds^2={4a^2\over(1+Z^2)^2}\big(Z^2d(t/a)^2-dZ^2\big)-
a^2\left({1-Z^2\over1+Z^2}\right)^2(d\th^2+\sin^2\th d\phi^2).
$$
$Z$ is related to the usual static radial coordinate $r$ by
$Z^2=(a-r)/(a+r)$, $a$ being the radius of the de Sitter sphere.

Calculation shows that, when written in terms of the local temperature $T$,
$\bra T^0_0\ket'$ takes the same form as for Rindler space. The
integrated quantity is then
$$
E'(\be,\ep,A)={\pi^2\over120a\be^4}\int^1_{\ep/2a}{(1-Z^2)^2\over Z^3}dZ
\int d\Omega\approx{A\pi^2\over60a\be^4\ep^2}
$$
and so the leading term in the entropy is
$$
S'(\be,\ep,A)\approx{A\pi^2\over45\ep^2\be^3}
$$
and
$$
S'(2\pi,\ep,A)\approx{A\over360\ep^2\pi}
$$
which again diverges as the ($Z$-radial) distance from the horizon $\ep$
tends to zero. In this case, $A=4\pi a^2$ is the actual area of the
horizon.

The two-dimensional black-hole obtained by removing the angular dependence
from the usual Schwarzschild solution is also easily treated as it is
conformally Rindler. $\bra T^0_0\ket'$ is again given by (\peq{2den}) but now
with $T^{-1}=4M\be(1-2M/r)^{1/2}$. Calculations similar to those already
detailed give the leading forms
$$
E'(\be,\ep)\approx{\pi\over24\be^2}\ln(D/\ep),\quad\quad
S'(\be,\ep)\approx{\pi\over3\be}\ln(D/\ep)
$$
and
$$
S'(2\pi,\ep)\approx{1\over6}\ln(D/\ep)
$$
as $\ep$, the $Z$-radial distance from the horizon, tends to zero.
Note that the variable corresponding to the Rindler $Z$ is here
dimensionless and equals $(r/2M-1)^{1/2}\ln(r/4M)$ in terms of the usual
radial coordinate.
\section{\bf Comments.}
We chose to open our discussion with the cosmic string results
but, of course, it is not necessary to go through these in order to obtain
the finite-temperature corrections. The method of [\pref{TGR}] is a direct
thermal calculation along the lines of Brown and Maclay's Casimir effect
analysis [\pref{Brown}]. The results take the particular form that they do
because of conformal relations between certain space-times, as discussed in
[\pref{Candelas}]. Actually, the detailed calculations of [\pref{TGR}]
involving coincidence limits can be bypassed by using the fact that
Rindler space-time is conformal to the open Einstein Universe and that the
heat-kernel expansion terminates on such a space. This leads onto the
topic of the global evaluation of thermodynamic quantities, which we briefly
mention, leaving a closer examination for a later communication.

We have obtained the total energy, entropy \etc by integrating the local
densities and the integration had to be stopped a distance $\ep$ from the
horizon. Regarding the divergence of the integrated quantities, it
is relevant to point
out that in the ordinary Casimir effect it can happen that, although the
local energy density diverges as the boundary is approached, a finite global
energy can be defined starting from a renormalised, or even a finite,
effective (one-loop) action. A possible way of reconciling these two total
energies is to introduce a divergent {\it boundary energy} to compensate for
the infinity caused by integrating the local density right up to the
boundary. The amplification of this statement involves a rather technical
discussion and we refer to [\pref{KCD,K}]. A  recent, relevant
implementation of this idea is [\pref{F}].

Maybe it is worth adding that it is possible to allow for a chemical potential,
and therefore a charged black hole, by a flux through the conical singularity.


\vskip 20truept
\noin{\bf{References}}
\vskip 10truept
\begin{putreferences}
\ref{Dowkera}{J.S.Dowker}
\ref{Callan}{C.Callan and F.Wilczek {\it On geometric entropy},
IASSNS-HEP-93/87, hep-th /9401072.}
\ref{Susskind}{L.Susskind and J.Uglum {\it Black hole entropy in canonical
quantum gravity and superstring theory}, Stanford preprint SU-TP-94-1,
hep-th/9401070.}
\ref{Kabat}{D.Kabat and M.J.Strassler {\it A comment on entropy and area},
Rutgers preprint RU-94-10, hep-th/9401125.}
\ref{Birrell}{N.D.Birrell and P.C.W.Davies {\it Quantum fields in curved
space}, Cambridge University Press 1982.}
\ref{t'Hooft}{t'Hooft \np{256}{85}{727}.}
\ref{condef}{J.S.Dowker {\it Quantum field theory around conical defects} in
{\it The formation and evolution of cosmic strings} edited by G.Gibbons,
S.W.Hawking and T.Vachas-\break pati, Cambridge University Press, 1990.}
\ref{qftcone}{J.S.Dowker \jpa{10}{77}{115}.}
\ref{Candelas}{P.Candelas and J.S.Dowker \pr{D19}{79}{2902}.}
\ref{Cand}{P.Candelas and D.Deutsch \prs{252}{77}{79}.}
\ref{TGR}{J.S.Dowker \pr{D18}{78}{1856}.}
\ref{Davies}{P.C.W.Davies \jpa{8}{75}{609}.}
\ref{Brown}{L.S.Brown and G.J.Maclay \pr{184}{69}{1272}.}
\ref{KCD}{G.Kennedy, R.Critchley and J.S.Dowker \anp{125}{80}{346}.}
\ref{K}{G.Kennedy \anp{138}{82}{353}.}
\ref{F}{D.V.Fursaev {\it The heat kernel expansion on a cone and quantum
fields near a cosmic string}, JINR preprint E2-93-291, hep-th/9309050.}
\ref{Takagi}{S.Takagi {\it Suppl. Prog. Theor. Phys.} {\bf 88} (1986) 1.}
\ref{Israel}{W.Israel \pl{A57}{76}{107}.}
\ref{Sewell}{G.L.Sewell \anp{141}{82}{201}.}
\ref{Fulling}{S.A.Fulling and S.N.N.Ruijsenhaars \prp{152}{87}{136}.}
\end{putreferences}
\bye